\documentclass[amsmath,11pt]{article}

\date{\empty}

\begin{document}

\title{\bf Large-scale magnetic fields in cosmology}

\author{Christos G. Tsagas\\ {\small Section of Astrophysics, Astronomy and Mechanics, Department of Physics}\\ {\small Aristotle University of Thessaloniki, Thessaloniki 54124, Greece}}


\maketitle

\begin{abstract}
Despite the widespread presence of magnetic fields, their origin, evolution and role are still not well understood. Primordial magnetism sounds appealing but is not problem free. The magnetic implications for the large-scale structure of the universe still remain an open issue. This paper outlines the advantages and shortcomings of early-time magnetogenesis and the typical role of $B$-fields in linear structure-formation scenarios.
\end{abstract}

\section{Introduction}\label{sI}
Magnetic fields are everywhere in the universe~\cite{K}. The Milky Way and many other galaxies possess fields of $\mu$Gauss strength, while ordered $B$-fields in the intracluster space suggest that structure formation could have been influenced by magnetic forces. In addition, Faraday rotation measurements at high redshifts indicate dynamically significant $B$-fields in protogalactic clouds. Despite its widespread presence, however, the origin of cosmic magnetism remains a mystery and a subject of debate. The alignment of the galactic fields supports the dynamo-amplification idea, but dynamos require an initial seed field to operate~\cite{P}. These seeds could be the result recent (post-recombination) physics, or have a primordial (pre-recombination) origin.\footnote{At recombination, roughly $10^5$~years after the Big Bang, matter ceases being ionised and the universe becomes `transparent' to photons. Relic of that epoch is the Cosmic Microwave Background (CMB) radiation~\cite{Pa}.} Deciding that is difficult, since the galactic fields have lost memory of their earlier history. In contrast, possible $B$-fields in the intercluster space, or magnetic imprints in the CMB should provide much better insight.

Primordial magnetism is attractive because it could explain all the magnetic fields seen in the universe, especially those in high redshift systems. Early magnetogenesis is not problem-free however~\cite{E}. Magnetic fields generated in the radiation era, namely between inflation and recombination, have too small coherence lengths and will distabilise the dynamo. Inflation can provide large-scale correlations, but $B$-fields that survived a epoch of de Sitter expansion are typically too weak to sustain the dynamo. Primordial magnetic fields must also comply with certain constraints. These come from nucleosynthesis and, mainly, the CMB, which limits the current magnetic strength between $10^{-6}$~Gauss (for random fields) and $10^{-9}$~Gauss (for homogeneous ones). Cosmological $B$-fields of $\mu$Gauus strength can affect structure formation, through their Lorentz force and the anisotropy that they introduce~\cite{TB}. The Lorentz force, in particular, generates and affects all types of density inhomogeneities. The anisotropic nature of the field, on the other hand, makes it a source of shear and gravitational waves.

\section{Cosmological electromagnetic fields}
Consider a general spacetime with metric $g_{ab}$, of signature ($-,+,+,+$), allow for a family of fundamental observers with 4-velocity $u_a$ (such that $u_au^a=-1$). The tensor $h_{ab}=g_{ab}+u_au_b$ projects orthogonal to $u_a$ into the observers' 3-D rest space. Together, $h_{ab}$ and $u_a$ introduce an 1+3 `threading' of the spacetime into time and space, while they decompose all physical and geometrical quantities into their timelike and spacelike parts. For instance, splitting the gradient of the 4-velocity,
\begin{equation}
\nabla_bu_a= {1\over3}\,\Theta h_{ab}+ \sigma_{ab}+ \omega_{ab}- A_au_a\,,  \label{Nbua}
\end{equation}
gives the kinematic variables~\cite{EvE}. These are the average volume expansion/contraction $\Theta=\nabla^au_a$, the shear $\sigma_{ab}={\rm D}_{\langle b}u_{a\rangle}$, the vorticity $\omega_{ab}={\rm D}_{[b}u_{a]}$ and the 4-acceleration $A_a=\dot{u}_a=u^b\nabla_bu_a$ ($\nabla_a$ is the covariant derivative operator and ${\rm D}_a=h_a{}^b\nabla_b$ is its 3-D counterpart).\footnote{Round brackets denote symmetrisation and square ones antisymmetrisation. Angled brackets indicate the symmetric and trace-free part of second-rank spacelike tensors and the orthogonally projected component of vectors. For example, ${\rm D}_{\langle b}u_{a\rangle}={\rm D}_{(b}u_{a)}-({\rm D}^cu_c)h_{ab}/3$ and $\dot{E}_{\langle a\rangle}=h_a{}^b\dot{E}_b$ -- see Eq.~(\ref{M1}). Also, overdots represent proper-time derivatives and primes conformal-time ones -- see Eq.~(\ref{cBwe}).} Similarly, the energy-momentum tensor of a general imperfect fluid, decomposes as~\cite{EvE}
\begin{equation}
T^{(m)}_{ab}= \rho u_au_b+ ph_{ab}+ 2q_{(a}u_{b)}+ \pi_{ab}\,,  \label{mTab}
\end{equation}
where $\rho$, $p$, $q_a$ and $\pi_{ab}$ are respectively the density, the isotropic pressure, the energy flux and the anisotropic pressure of the matter. Also, $q_au^a=0=\pi_{ab}u^b$, with $\pi_{ab}=\pi_{ba}$ and $\pi_a{}^a=0$. For a perfect fluid $q_a=0=\pi_{ab}$ and for a barotropic medium $p=p(\rho)$. Note that we use geometrised units, with $c=1=8\pi G$, throughout this paper.

Relative to the fundamental observers, the electromagnetic field splits into an electric and a magnetic component, represented by the vectors $E_a$ and $B_a$ respectively (with $E_au^a=0=B_au^a$). Then, the stress-energy tensor of the Maxwell field reads as~\cite{TB}
\begin{equation}
T^{(em)}_{ab}= {1\over2}\,\left(E^2+B^2\right)u_au_b+ {1\over6}\,\left(E^2+B^2\right)h_{ab}+ 2\mathcal{Q}_{(a}u_{b)}+ \mathcal{P}_{ab}\,,  \label{emTab}
\end{equation}
with $E^2=E_aE^a$, $B^2=B_aB^a$, $\mathcal{Q}_a= \varepsilon_{abc}E^bB^c$ being the Poynting vector and $\mathcal{P}_{ab}=-E_{\langle a}E_{b\rangle}-B_{\langle a}B_{b\rangle}$ the anisotropic electromagnetic pressure. By construction $\mathcal{Q}_au^a=0=\mathcal{P}_{ab}u^a$, with $\mathcal{P}_{ab}=\mathcal{P}_{ba}$ and $\mathcal{P}_a{}^a=0$.\footnote{The totally antisymmetric tensor $\varepsilon_{abc}$, with $\varepsilon_{abc}u^c=0$, is the Levi-Civita symbol of the 3-D space.} In other words, the Maxwell field corresponds to an imperfect fluid with $\rho^{(em)}=(E^2+B^2)/2$, $p^{(em)}=(E^2+B^2)/6$, $q^{(em)}_a=\mathcal{Q}_a$ and $\pi^{(em)}_{ab}=\mathcal{P}_{ab}$.

The evolution of the electromagnetic field is monitored by Maxwell's equations. In the $u_a$-frame, these decompose into a set of two propagation equations~\cite{TB},
\begin{equation}
\dot{E}_{\langle a\rangle}= \left(\sigma_{ab}+\omega_{ab}- {2\over3}\,\Theta h_{ab}\right)E^b+ \varepsilon_{abc}A^bB^c+ {\rm curl}B_a- \mathcal{J}_a\,,  \label{M1}
\end{equation}
\begin{equation}
\dot{B}_{\langle a\rangle}= \left(\sigma_{ab}+\omega_{ab}- {2\over3}\,\Theta h_{ab}\right)B^b- \varepsilon_{abc}A^bE^c- {\rm curl}E_a\,,  \label{M2}
\end{equation}
which are supplemented by the constraints
\begin{equation}
{\rm D}^aE_a= \mu- 2\omega^aB_a \hspace{15mm} {\rm and} \hspace{15mm} {\rm D}^aB_a= 2\omega^aE_a\,.  \label{M4}
\end{equation}
Here $\mathcal{J}_a$ is the electric 3-current, $\mu$ is the electric charge and ${\rm curl}B_a=\varepsilon_{abc}{\rm D}^bB^c$  by definition (with an analogous expression for ${\rm curl}E_a$). The 3-current is related to the electric field via Ohm's law. For a single charged fluid the latter takes the form
\begin{equation}
\mathcal{J}_a=\varsigma E_a\,,  \label{Ohm}
\end{equation}
with $\varsigma$ representing the electrical conductivity of the medium~\cite{G}. At the ideal magnetohydrodynamic (MHD) limit the conductivity is very high, with $\varsigma\rightarrow\infty$. Then, the electric field vanishes and the currents keep the magnetic field frozen-in with the fluid. At the opposite end, where $\varsigma\rightarrow0$, the currents are zero despite the presence of a finite electric field. Using the set (\ref{emTab})-(\ref{Ohm}) we can follow the evolution of electromagnetic fields in a variety of cosmological environments and study the implications of large-scale $B$-fields, in particular, for the formation and the evolution of the structure that we see in our universe today.

\section{Primordial magnetic fields}
The detection of coherent magnetic fields in remote astrophysical systems increases the possibility of significant $B$-fields of cosmological origin. Early magnetogenesis, however, faces problems with both the strength and the size of the primordial field. Typically, magnetic fields generated between inflation and recombination have too small coherence lengths and cannot seed the galactic dynamo. The reason is causality, which limits the size of the seed to the horizon scale at the time of magnetogenesis. This is typically much smaller than the 10~Kpc length required by the dynamo. A mechanism known as `inverse cascade' can increase the correlation length by transferring magnetic energy to larger scales, but needs large-amounts of helicity~\cite{B}.\footnote{See also Prof.~Brandenburg's contribution to this issue.}

Inflation has long been seen as a solution to the scale problem, since it naturally creates super-horizon correlations. There is a serious strength problem however. Magnetic fields that survive an epoch of typical inflationary expansion are weaker than $10^{-50}$~G. This lies well below the dynamo requirements, which vary between $10^{-12}$ and $10^{-34}$~Gauss, in today's values, depending on the efficiency of the amplification and the cosmological model it operates in~\cite{P,TB}. This dramatic depletion is attributed to the `adiabatic decay' of large-scale magnetic fields, which typically dilute as $a^{-2}$ ($a$ is the cosmological scale factor). To get an idea why, consider the expressions (\ref{M1}) and (\ref{M2}) of the previous section. These combine to give a wave equation for each component of the Maxwell field~\cite{T1}. On a Friedmann-Robertson-Walker (FRW) background, the linear wave formula for the $n$-th magnetic mode reads as
\begin{equation}
\mathcal{B}^{\prime\prime}_{(n)}+ n^2\mathcal{B}_{(n)}= -2K\mathcal{B}_{(n)}\,,  \label{cBwe}
\end{equation}
where $\mathcal{B}_{(n)}=a^2B_{(n)}$ is the rescaled magnetic vector, $n$ is the eigenvalue of the mode, $K=0,\pm1$ is the 3-curvature index of the FRW background and the primes indicate conformal-time derivatives.\footnote{We use continuous harmonic eigenvalues, with $n^2\geq0$, in spatially flat and open FRW models and discrete ones, with $n^2\geq3$, when $K=+1$~\cite{T1}.} When $K=0$ the above reduces to the Minkowski-like expression
\begin{equation}
\mathcal{B}^{\prime\prime}_{(n)}+ n^2\mathcal{B}_{(n)}= 0\,,  \label{fcBwe}
\end{equation}
which solves to give $B_{(n)}\propto a^{-2}$ at all times and on all scales. In practice, this result translates into a current (comoving) strength below $10^{-50}$~G for typical `inflationary' magnetic fields. Therefore, the adiabatic ($B_{(n)}\propto a^{-2}$) decay needs to slow down if we are to have inflation-generated $B$-fields of astrophysical relevance. The effect is known as `superadiabatic amplification' and it is usually achieved outside classical electrodynamics (e.g.~see~\cite{TW} for a representative though incomplete list), unless FRW models with nonzero spatial geometry are employed~\cite{TK}.\footnote{Friedmann models with non-Euclidean spatial geometry are only locally conformal to Minkowski space. On these backgrounds the adiabatic ($B\propto a^{-2}$) magnetic decay is guaranteed only on scales well inside the curvature radius. Near and beyond the curvature length, the magneto-geometrical term in the right-hand side of Eq.~(\ref{cBwe}) becomes important and can change the standard evolution of the $B$-field.}

\section{Magnetic effects on structure formation}
The isotropy of the CMB strongly suggests that our universe was extremely smooth at recombination. Today, however, we see structure all around us and the obvious question is how this structure was formed. Gravitational instability seems to hold the answer, but we do not know the details yet. The $\Lambda$CDM model is the current `concordance' scenario, which however excludes magnetic fields and has a rather large number of free parameters.\footnote{The $\Lambda$CDM model assumes an accelerating universe dominated primarily by a cosmological constant ($\Lambda$) -- or by dynamical dark energy acting as an effective cosmological constant -- and secondarily by Cold Dark Matter (CDM). Baryons make only a small fraction ($\sim5\%$) of the total matter.}

Studies of magnetised structure formation typically work within the ideal MHD approximation and look at the effects of the magnetic Lorentz force on density inhomogeneities.\footnote{Beyond the linear regime, one should also account for microphysical processes and departures from the ideal MHD limit. As yet, however, studies of this nature are both sparse and patchy.} These generally come in the form of scalar, vector and (trace-free) tensor distortions. The former are those commonly known as density perturbations and represent overdensities or underdensities in the matter distribution. Vector inhomogeneities are monitored by the curl of the density gradient and describe rotational, vortex-like, density perturbations. Finally, tensor inhomogeneities correspond to shape distortions. Following~\cite{TB}, the scalar
\begin{equation}
\Delta= {a^2\over\rho}\,{\rm D}^2\rho\,,  \label{divs}
\end{equation}
describes linear density perturbations (${\rm D}^2={\rm D}^a{\rm D}_a$ is the 3-D Laplacian operator). In a perturbed, weakly magnetised, spatially flat FRW universe, the above evolves according to~\cite{TB}
\begin{equation}
\dot{\Delta}= 3wH\Delta- (1+w)\mathcal{Z}+ {3\over2}\,c_{\rm a}^2(1+w)H\mathcal{B}\,,  \label{dotDelta}
\end{equation}
where $\mathcal{Z}=a^2{\rm D}^2\Theta$ and $\mathcal{B}= (a^2/B^2){\rm D}^2 B^2$ describe linear inhomogeneities in the expansion and the magnetic energy density respectively.\footnote{Equation (\ref{dotDelta}) shows that $B$-fields are generic sources of linear density perturbations. Indeed, even when $\Delta$ and $\mathcal{Z}$ are zero initially, $\dot{\Delta}$ will take nonzero values solely due to the magnetic presence. Note that only the pressure part of the Lorentz force contributes to the linear expressions (\ref{dotDelta}) and (\ref{dotcZ}). To account for the tension effects, one needs to allow for FRW backgrounds with non-Euclidean spatial geometry~\cite{TB}.} To first order, these two variables propagate as
\begin{eqnarray}
\dot{\mathcal{Z}}&=& -2H\mathcal{Z}- {1\over2}\,\rho\Delta+ {1\over4}\,c_{\rm a}^2(1+w)\rho\mathcal{B}- {c_s^2\over1+w}{\rm D}^2\Delta- {1\over2}\,c_{\rm a}^2{\rm D}^2\mathcal{B}  \label{dotcZ}
\end{eqnarray}
and
\begin{equation}
\dot{\mathcal{B}}= {4\over3(1+w)}\,\dot{\Delta}+ {4(c_s^2-w)H\over1+w}\,\Delta\,,  \label{dotcB}
\end{equation}
respectively. Also, $w=p/\rho$, $H=\dot{a}/a$ is the background Hubble parameter, $c_s^2=\dot{p}/\dot{\rho}$ is the square of the adiabatic sound speed and $c_{\rm a}^2=B^2/\rho(1+w)$ is that of the Alfv\'en speed. Finally, we note that $B^2\ll\rho$, given the relative weakness of the magnetic field.

The system (\ref{dotDelta})-(\ref{dotcB}) has analytical solutions in the radiation and the dust eras~\cite{TB}. Before equipartition, when $w=1/3=c_s^2$, $H=1/2t$, $\rho=3/4t^2$ and $c_{\rm a}^2=3B^2/4\rho=\,$constant, large-scale magnetised density perturbations obey the power-law solution
\begin{equation}
\Delta= \mathcal{C}_0+ \mathcal{C}_1t^{-{1\over2}+{10\over9}\,c_{\rm a}^2}+ \mathcal{C}_2t^{-{4\over3}\,c_{\rm a}^2}+ \mathcal{C}_3t^{1-{4\over9}\,c_{\rm a}^2}\,.  \label{lsrDelta}
\end{equation}
In the absence of the $B$-field we recover the standard growing mode of $\Delta\propto t$, which means that the magnetic presence reduces the growth rate by $4c_{\rm a}^2/9$. Well inside the horizon, the $k$-mode of the density contrast oscillates like a magneto-sonic wave with
\begin{equation}
\Delta_{(k)}\propto \sin\left[c_s\left(1+{2\over3}\,c_{\rm a}^2\right) \left({\lambda_H\over\lambda_k}\right)_0\sqrt{t\over t_0}\, \right]\,,  \label{ssrDelta}
\end{equation}
where $\lambda_k=a/k$ is the perturbed scale and $\lambda_H=1/H$ the Hubble horizon~\cite{TB}. Here, the magnetic pressure increases the effective sound speed and therefore the oscillation frequency. The former makes the Jeans length larger than in non-magnetised models. The latter brings the peaks of short-wavelength oscillations in the radiation density closer, leaving a potentially observable signature in the CMB spectrum~\cite{ADGR}.

When dust dominates, $w=0=c_s^2$, $H=2/3t$, $\rho=4/3t^2$ and $c_{\rm a}^2=B^2/\rho\propto t^{-2/3}$. Then, on superhorizon scales, density perturbations evolve as~\cite{TB}
\begin{equation}
\Delta= \mathcal{C}_0+ \mathcal{C}_1t^{-2/3}+ \mathcal{C}_2t^{\alpha_1}+ \mathcal{C}_3t^{\alpha_2}\,,
\label{dDelta}
\end{equation}
with $\alpha_{1,2}= -[1\pm5\sqrt{1-(32/75) (c_{\rm a}\,\lambda_H/\lambda_k)_0^2}]/6$. In the absence of the $B$-field we recover again the standard solution with $\alpha_1=2/3$ and $\alpha_2=-1$. As before, the magnetic presence slows down the growth rate of density perturbations. Also, the field's pressure leads to a magnetically induced Jeans length, below which density perturbations cannot grow. The magnetic Jeans scale, as a fraction of the Hubble radius, is~\cite{TB}
\begin{equation}
\lambda_J\sim c_{\rm a}\lambda_H\,.  \label{mJeans}
\end{equation}
Assuming a $B$-field of approximately $10^{-9}$~G, we find that $\lambda_J\sim10$~Kpc. Alternative, magnetic fields close to $10^{-7}$~G, like those found in galaxies and galaxy clusters, give $\lambda_J\sim1$~Mpc. This lies intriguingly close to the size of a cluster of galaxies. Overall, the magnetic effect on density perturbations seems rather negative. Although $B$-fields generate this type of distortions, they do not help them to grow. Instead, the magnetic presence either suppresses the growth rate of density perturbations or increases the effective Jeans length and therefore the domain where these inhomogeneities cannot grow.

Magnetic fields also induce and affect rotational, vortex-like, density inhomogeneities. These are described by the vector $\mathcal{W}_a=-(a^2/2\rho)\varepsilon_{abc}{\rm D}^b{\rm D}^c\rho$, which on an FRW background and after matter-radiation equality evolves as
\begin{equation}
\ddot{\mathcal{W}}_a= -4H\dot{\mathcal{W}}_a- {1\over2}\,\rho\mathcal{W}_a+ {1\over3}\,c_{\rm a}^2{\rm D}^2\mathcal{W}_a\,,  \label{ddotW}
\end{equation}
to linear order~\cite{TB}. Defining $\lambda_{\rm a}=c_{\rm a}\lambda_H$ as the `Alfv\'en horizon', we may write the associated solution in the form~\cite{TB}
\begin{equation}
\mathcal{W}_{(k)}= \mathcal{C}_1t^{\alpha_1}+ \mathcal{C}_2t^{\alpha_2}\,,  \label{dW}
\end{equation}
with $\alpha_{1,2}= -[5\pm\sqrt{1-(48/9) (\lambda_{\rm a}/\lambda_k)^2_0}]/6$. On scales far exceeding the Alfv\'en horizon, $\lambda_{\rm a}\ll\lambda_k$ and the perturbed mode decays as $\mathcal{W}\propto t^{-2/3}$. This rate is considerably slower than $\mathcal{W}\propto t^{-1}$, the decay rate associated with magnetic-free dust cosmologies. Well inside $\lambda_{\rm a}$, on the other hand, magnetised vortices oscillate like Alfv\'en waves, with~\cite{TB}
\begin{equation}
\mathcal{W}_{(k)}\propto t^{-5/6}\cos\left[{2\sqrt{3}\over9} \left({\lambda_{\rm a}\over\lambda_k}\right)_0\ln t\right]\,.  \label{sscW}
\end{equation}
Thus, the effect of the $B$-field on a given vortex mode is to reduce its standard depletion rate. Analogous is the magnetic effect on $\omega_a$, the vorticity proper. Overall, magnetised cosmologies rotate faster than their magnetic-free counterparts. In contrast to density perturbations, the field seems to favour the presence of vorticity. This qualitative difference should probably be attributed to the fact that the tension part of the Lorentz force also contributes to Eq.~(\ref{ddotW}).

\section{Discussion}
In addition to scalar and vector perturbations, magnetic fields also generate and affect tensor-type inhomogeneities that describe shape-distortions in the density distribution~\cite{TB}. An initially spherically symmetric inhomogeneity, for example, will change shape due to the magnetically induced anisotropy. All these are the effects of the Lorentz force. Even when the latter is removed from the system, however, the $B$-field remains active. Due to its energy density and anisotropic nature, for example, magnetism affects both the local and the long-range gravitational field. The anisotropic magnetic pressure, in particular, leads to shear distortions and subsequently to gravitational-wave production. Overall, magnetic fields are a very versatile source. They are also rather unique in nature, since $B$-fields are the only known vector source of energy. An additional unique magnetic feature, which remains relatively unexplored, is its tension. When we add to all these the widespread presence of magnetic fields, it makes sense to say that no realistic structure formation scenario should a priori exclude them. It was probably thoughts like this that motivated some researchers to start adding magnetic fields into their numerical codes~\cite{DSGF}. Hopefully their numbers will increase and we will soon have structure formation models with fewer free parameters and more physics.

\end{document}